\documentclass[twocolumn,showpacs,aps,prl,letterpaper]{revtex4}
\usepackage{graphicx}  
\usepackage{dcolumn}   
\usepackage{bm}        
\usepackage{amssymb}   
\usepackage{slashed}   
\usepackage{amsmath}   
\long\def\inst#1{\par\nobreak\kern 4pt\nobreak
    {\itshape #1}\par\vskip 10pt plus 3pt minus 3pt}

\begin{document}



\title{Radiative Leptonic Decays of the charged $B$ and $D$ Mesons Including Long-Distance Contribution}
\author{Ji-Chong Yang}
\author{Mao-Zhi Yang}
\affiliation{School of Physics, Nankai University, Tianjin 300071, P.R. China}
\vskip 0.25cm

\date{\today}

\begin{abstract}
In this work we study the radiative leptonic decays of $B^-$, $D^-$ and $D_s^-\to \gamma l \bar{\nu}$, including both the short-distance and long-distance contributions. The short-distance contribution is calculated by using the relativistic quark model, where the bound state wave function we used is that obtained in the relativistic potential model. The long-distance contribution is estimated by using vector meson dominance model.
\end{abstract}

\pacs{13.20.Fc, 13.20.He, 12.39.Ki, 12.40.Vv}

\maketitle

\section{\label{sec:level1}I Introduction}

The mechanism of heavy meson decays is one of the most interesting and challenging fields in particle physics, it involves both strong and weak interactions. Nowadays strong interaction in the non-perturbative region is still an unsolved problem. Compared with hadronic decays, leptonic decay is simpler. Strong interaction only occurs within the initial particle. Pure-leptonic decay of heavy meson can be used to determine the decay constant, which describes the possibility-amplitude for the quark-untiquark emerging at the same point. The pure-leptonic decay is helicity suppressed. The decay branching ratio of a pseudoscalar meson $P$ with quark content $\bar{Q}q$ within the standard model is
\begin{equation}
B(P\to l\bar{\nu})=\frac{G_F^2|V_{Qq}|^2}{8\pi}\tau_Pf_P^2m_l^2m_P(1-\frac{m_l^2}{m_P^2})^2,
\end{equation}
where $G_F$ is the Fermi coupling constant, $V_{Qq}$ the Cabibbo-Kobayashi-Maskawa (CKM) matrix element, $\tau_P$ the life time of the meson $P$, and $m_P$ and $m_l$ the masses of the meson $P$ and lepton $l$, respectively. The decay rate is proportional to the lepton mass squared $m_l^2$ is the consequence of the helicity suppression. However, the presence of one photon in the final state can compensate the helicity suppression. As a result, the radiative leptonic decay can be as large as, or even larger than the pure-leptonic decay mode. It thus opens a window for detecting the dynamics of strong interaction in the heavy meson or studying the effect of strong interaction in the decay.

The radiative leptonic decay rates of the charged $B$ and $D$ mesons have been studied with various methods in the literature. In Ref. \cite{workbefore11}, $B$ and $D_s\to l\bar{\nu}\gamma $ are calculated in a non-relativistic quark model, the branching ratios of the order of $10^{-4}$ for $D_s\to l\bar{\nu}\gamma$ and $10^{-6}$ for $B\to l\bar{\nu}\gamma$ are found. In Ref. \cite{workbefore12} with perturbative QCD approach, it is found that the branching ratio of $D_{s}^+\to e^+\nu\gamma$ is of the order of $10^{-3}$ and $D^{+}\to e^+\nu\gamma$ of the order of $10^{-4}$, while the branching ratio of $B^{+}\to e^+\nu\gamma$ is at the order of $10^{-6}$. On the other hand, a smaller branching ratio is obtained for $D_{(s)}\to l\bar{\nu}\gamma $ within the light front quark model \cite{workbefore13}. Smaller result for $D_{(s)}\to l\bar{\nu}\gamma$ is also obtained in Ref. \cite{workbefore2} within the non-relativistic constituent quark model, which gives that the branching ratio of $D^{-}\to l\bar{\nu}\gamma$ is of the order of $10^{-6}$ and $D^{-}_s\to l\bar{\nu}\gamma$ of the order of $10^{-5}$. The problem of factorization in QCD for $B\to l\nu\gamma$ is studied in Ref. \cite{Sachrajda}.

In this work, we study the radiative leptonic decays of the charged $B$, $D$ and $D_s$ mesons  to $l\bar{\nu}\gamma$ including both the short and long-distance contributions. The short-distance contribution is considered at tree level. The wave function of the heavy meson used here is obtained in the relativistic potential model previously \cite{sdbase}. The long-distance contribution is estimated by using the idea of the vector meson dominance (VMD) \cite{VMD1,VMD2,VMD3,VMD4,VMD5} followed by the transition of the vector meson to a photon. We find that the long-distance contribution can enhance the decay rates seriously.

The remaining part of this paper is organized as follows. In Sec.II, we present the short-distance amplitude. In Sec.III, the long-distance contribution is considered. The numerical results and discussion are given in Sec.IV. Sec.V is a brief summary.

\subsection{\label{sec:level2}II The Short-Distance Contribution}

We use $P$ to denote the pseudoscalar meson which is composed of a heavy anti-quark $\bar{Q}$ and a light quark $q$, such as $B$ and $D$ mesons. There are four Feynman diagrams contributing to the radiative decays $P^-\to l \bar{\nu}\gamma$ at tree level, which are shown in Fig.~\ref{fig:feynsd}. However the contribution of Fig.~\ref{fig:feynsd} (d) is suppressed by a factor of $1\left/M_w^2\right.$, it can be neglected for simplicity. The effective Hamiltonians corresponding to the other three diagrams in Fig.~\ref{fig:feynsd} can be written as:
\begin{figure}
\includegraphics[scale=0.6]{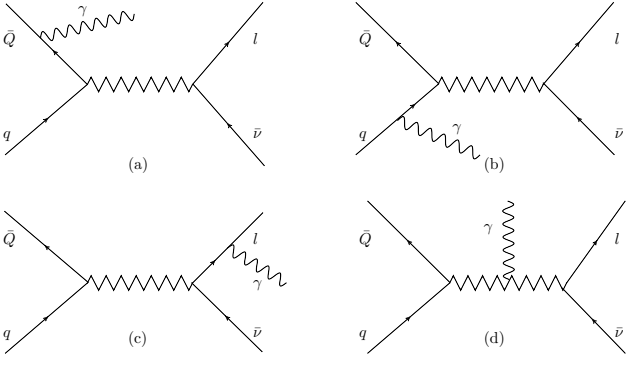}
\caption{\label{fig:feynsd} Feynman diagrams of short-distance contribution at tree level.}
\end{figure}

\begin{equation}
\begin{split}
&\mathcal{H}_a=\frac{- i e G_F V_{Qq}}{\sqrt{2}}Q_Q\bar{Q}\slashed A \frac{\slashed p_{\gamma }-\slashed p_Q+m_Q}{2\left(p_{\gamma }{\cdot}p_Q\right)}P_L^{\mu } q\left(lP_{L\mu}\bar{\nu}\right),\\
&\mathcal{H}_b=\frac{- i e G_F V_{Qq}}{\sqrt{2}}Q_q\bar{Q}P_L^{\mu }
\frac{\slashed p_q-\slashed p_{\gamma }+m_q}{2\left(p_{\gamma }{\cdot}p_q\right)}\slashed A q\left(lP_{L\mu}\bar{\nu}\right),\\
&\mathcal{H}_c=\frac{- i e G_F V_{Qq}}{\sqrt{2}}Q_Q\bar{Q}P_L^{\mu } q\left(l\slashed A\frac{\slashed p_{\gamma }+\slashed p_l+m_l}{2\left(p_{\gamma }{\cdot}p_l\right)}P_{L\mu}\bar{\nu}\right)
,\label{eq:one}
\end{split}
\end{equation}
where $P_L^{\mu }$ is defined as $\gamma ^{\mu }\left(1-\gamma _5\right)$, and $V_{Qq}$ represents for the CKM matrix elements. $Q_Q$ and $Q_q$ are the electric charges of the quarks  $Q$ and $q$, respectively. $A$ is the electro-magnetic field.

$\mathcal{H}_a$, $\mathcal{H}_b$ and $\mathcal{H}_c$ can be divided into two terms for convenience, according to the numerator of the fermion propagator. For example, $\mathcal{H}_a$ can be written as
\begin{equation}
\mathcal{H}_a=\frac{- i e Q_Q G_F V_{Qq}}{2\sqrt{2}}\left(\mathcal{M}_{\text {a1}}+2\mathcal{M}_{\text {a2}}\right)^{\mu}\left(lP_{\text {L$\mu $}}\bar{\nu}\right)
\label{eq:two},
\end{equation}
where
\begin{equation}
\begin{split}
\mathcal{M}_{a1}^{\mu}&=\bar{Q}\slashed A\frac{\slashed p_{\gamma }}{p_{\gamma }{\cdot}p_Q}\gamma ^{\mu }\left(1-\gamma _5\right) q.\\
&=- i \epsilon ^{\alpha \beta \mu \sigma }A_{\alpha}p_{\gamma\beta}v_{a\sigma}+p_{\gamma }^{\mu }\left(A{\cdot}v_a\right)-A^{\mu}\left(p_{\gamma }{\cdot}v_a\right),\\
\mathcal{M}_{a2}^{\mu}&=\frac{1}{2}\bar{Q}\slashed A\frac{m_Q-\slashed p_Q}{p_{\gamma }{\cdot}p_Q}\gamma ^{\mu }\left(1-\gamma _5\right) q = - A _{\alpha} t_a^{\alpha \mu },
\label{eq:three}
\end{split}
\end{equation}
with
\begin{equation}
\begin{split}
&v_a^{\mu }=\bar{Q}\frac{1}{p_{\gamma }{\cdot}p_Q}\gamma ^{\mu }\left(1-\gamma _5\right) q,\\
&t_a^{\alpha \mu }=\bar{Q}\frac{p_Q^{\alpha }}{p_Q{\cdot}p_{\gamma }}\gamma ^{\mu}\left(1-\gamma _5\right)q
\label{eq:four}.
\end{split}
\end{equation}
The amplitude of the radiative leptonic decay can be obtained by inserting the operator of the effective Hamiltonian between the initial and final particle states. For example, the contribution of Fig.~\ref{fig:feynsd} (a) is
\begin{equation}
\begin{split}
\mathcal{A}_a&=<\gamma \bar{\nu} l\mid\mathcal{H}_a\mid P>\\
&= \frac{- i e Q_Q G_F V_{Qq}}{2\sqrt{2}}  \left(u_l P_{\text {L$\mu $}}v_{\bar{\nu}}\right)\\
&\times \left(<\gamma \mid \mathcal{M}_{\text {a1}} \mid P>+2 <\gamma \mid \mathcal{M}_{a2} \mid P>\right) \\
&= \frac{- i e Q_Q G_F V_{Qq}}{2\sqrt{2}}  \left(u_l P_{\text {L$\mu $}}v_{\bar{\nu}}\right)\\
&\times \left(- i \epsilon ^{\alpha \beta \mu \sigma }\varepsilon _{\gamma\alpha}^*p_{\gamma\beta}<0\mid v_{a\sigma}\mid P>+p_{\gamma }^{\mu }\varepsilon _{\gamma }^*{\cdot}<0\mid v_a\mid P>\right.\\
&\left.-\varepsilon _{\gamma }^{\mu *}p_{\gamma }{\cdot}<0\mid v_a\mid P> -2 \varepsilon _{\gamma \alpha}^*<0\mid t_a^{\alpha \mu }\mid P>\right)
\label{eq:five}.
\end{split}
\end{equation}

The matrix elements $<0\mid v_a^{\mu }\mid P>$ and $<0\mid t_a^{\alpha \mu }\mid P>$ only depend on the momenta $p_P$ and $p_{\gamma}$. According to their Lorentz structure, they can be decomposed as a linear combination of two terms of $p_P$ and $p_{\gamma}$
\begin{equation}
\begin{split}
V_a^{\mu}&=<0 \mid \bar{Q}\frac{1}{p_{\gamma }{\cdot}p_Q}\gamma ^{\mu }\left(1-\gamma _5\right) q \mid P>\\
&=\frac{m_P}{p_P{\cdot}p_{\gamma }}\left(A_{a1}p_P^{\mu }+B_{a1}\frac{m_P^2}{p_P{\cdot}p_{\gamma }}p_{\gamma }^{\mu }\right)\\
T_a^{\alpha \mu }&=<0 \mid \bar{Q}\frac{p_Q^{\alpha }}{p_Q{\cdot}p_{\gamma }}\gamma ^{\mu}\left(1-\gamma _5\right)q \mid P>\\
&= \frac{m_P}{p_P{\cdot}p_{\gamma }}\left[A_{a2}p_P^{\alpha }p_P^{\mu }+\frac{m_P^2}{p_P{\cdot}p_{\gamma }}\left(B_{a2}p_P^{\alpha }p_{\gamma }^{\mu }+C_{a2}p_P^{\mu }p_{\gamma }^{\alpha }\right)\right. \\ & \left.
+\frac{m_P^4}{\left(p_P{\cdot}p_{\gamma }\right)^2}D_{a2}p_{\gamma }^{\alpha }p_{\gamma }^{\mu }+E_{a2}m_P^2g^{\alpha \mu }\right. \\
& \left. +F_{a2}\frac{m_P^2}{p_P{\cdot}p_{\gamma }}\epsilon ^{\alpha \mu \rho \sigma}p_{P\rho }p_{\gamma \sigma }\right]
\label{eq:six}.
\end{split}
\end{equation}
The coefficients $A_{a1}$, $B_{a1}$, $A_{a2}$, $B_{a2}$, $C_{a2}$, $D_{a2}$, $E_{a2}$ and $F_{b2}$ are all dimensionless constants. The terms of $B_{a1}$, $C_{a2}$ and $D_{a2}$ do not contribute to the decay amplitude $\mathcal{A}_a$ when substituting the above decomposition into eq.\ref{eq:five}. Therefore these terms can be dropped. The coefficients can be obtained by the treatment in the following. Multiplying $V_a^{\mu}$ with $p_{\gamma \mu}$, we can obtain $A_{a1}$ as
\begin{equation}
A_{a1}=\frac{1}{m_P}<0\mid \bar{Q}\frac{1}{p_{\gamma }{\cdot}p_Q}\slashed p_{\gamma}\left(1-\gamma _5\right) q \mid P>
\label{eq:seven}.
\end{equation}
Similarly, multiplying $T_a^{\alpha \mu }$ with $p_{\gamma \alpha}$, we have
\begin{equation}
\begin{split}
&B_{a2}+E_{a2}=0,\\
&A_{a2}=\frac{1}{m_P^3}<0 \mid \bar{Q}\slashed p_P\left(1-{\gamma }_5\right)q\mid P>
\label{eq:eight}.
\end{split}
\end{equation}
Multiplying $T_a^{\alpha \mu }$ with $p_{\gamma \mu}p_{P\alpha}$ and $g_{\alpha\mu}$, and using $B_{a2}=-E_{a2}$, one can get
\begin{equation}
\begin{split}
&E_{a2}=-B_{a2}
=\frac{1}{2 m_P^3}<0\mid \bar{Q}\left[\left(p_P\cdot p_{\gamma}\right)\slashed p_Q \right.\\
&\left.- \left(p_P\cdot p_Q\right)\slashed p_{\gamma} \right]\frac{\left(1-{\gamma }_5\right)}{p_Q \cdot p_{\gamma}}q\mid P>
\label{eq:nine}.
\end{split}
\end{equation}
Finally $F_{a2}$ can obtained by multiplying $T_a^{\alpha \mu }$ with $\epsilon_{\alpha \mu \beta \nu}p_{\gamma}^{\nu}p_P^{\beta}$:
\begin{equation}
\begin{split}
F_{a2}=\frac{p_{\gamma}^{\nu}p_P^{\beta}}{2 m_P^3}<0\mid \bar{Q}\frac{\epsilon _{\alpha\mu\beta\nu}p_Q^{\alpha}}{p_Q\cdot p_{\gamma}}\gamma ^{\mu}\left(1-\gamma _5\right)q \mid P>
\label{eq:ten}.
\end{split}
\end{equation}
The amplitude $\mathcal{A}_b$ can be treated in the same way with some coefficients defined as follows
\begin{equation}
\begin{split}
&<0\mid\bar{Q}\frac{1}{p_{\gamma }\cdot p_q}\gamma ^{\mu }\left(1-\gamma _5\right) q\mid P>=\\
&\frac{m_P}{p_P\cdot p_{\gamma }}\left(A_{\text {b1}}p_P^{\mu }+B_{\text {b1}}\frac{m_P^2}{p_P\cdot p_{\gamma }}p_{\gamma }^{\mu }\right),\\
&<0\mid \bar{Q}\frac{p_q^{\alpha}}{p_{\gamma}{\cdot}p_q}
{\gamma}^{\mu}\left(1-{\gamma}_5\right)q\mid P>=\\
&\frac{m_P}{p_P{\cdot}p_{\gamma}}
\left[A_{b2}p_P^{\alpha}p_P^{\mu}+
\frac{m_P^2}{p_P{\cdot}p_{\gamma}}
\left(B_{b2}p_P^{\alpha}p_{\gamma}^{\mu}+C_{b2}p_P^{\mu}p_{\gamma}^{\alpha}\right)+\right.\\
&\left.\frac{m_P^4}{\left(p_P{\cdot}p_{\gamma}\right)^2}D_{b2}p_{\gamma}^{\alpha}p_{\gamma}^{\mu}+
E_{b2}m_P^2 g^{\alpha\mu}\right]
\label{eq:11}.
\end{split}
\end{equation}
Using the matrix element $A_{a2} m_P p_P^{\mu}=<0\mid \bar{Q} {\gamma}^{\mu}\left(1-\gamma _5\right) q \mid P>$, the amplitude $\mathcal{A}_c$ can be treated simlarly. Finally, the total amplitude can be expressed as
\begin{equation}
\begin{split}
&\mathcal{A}_a =\frac{- i e G_F V_{Qq}}{2\sqrt{2}}\left(u_l P_{\text {L$\mu $}}v_{\bar{\nu}}\right)\left\{ -\frac{Q_Qm_PA_{a1}}{p_P{\cdot}p_{\gamma}}i \epsilon ^{\alpha \beta \mu \sigma}p_{\gamma \beta }\varepsilon ^*_{\gamma \alpha}p_{P\sigma}\right.\\
&\left.-\left[\frac{Q_Q m_P A_{a1}}{p_P\cdot p_{\gamma}}+\frac{2 Q_Q E_{a2} m_P^3}{\left(p_P\cdot p_{\gamma}\right)^2}\right]\left[\left(p_P{\cdot}p_{\gamma}\right)\varepsilon _{\gamma}^{{\mu}*}-\left(p_P{\cdot}\varepsilon\right)p_{\gamma}^{\mu}\right]
\right.\\
&\left.-2\left(Q_Q\frac{p_P{\cdot}\varepsilon}{p_P{\cdot}p_{\gamma}}\right)A_{a2}m_P p_P^{\mu}\right\},\\
&\mathcal{A}_b =\frac{- i e G_F V_{Qq}}{2\sqrt{2}}\left(u_l P_{\text {L$\mu $}}v_{\bar{\nu}}\right)\left\{ -\frac{Q_qm_PA_{b1}}{p_P{\cdot}p_{\gamma}}i \epsilon ^{\alpha \beta \mu \sigma}p_{\gamma \beta }\varepsilon ^*_{\gamma \alpha}p_{P\sigma}\right.\\
&\left.+\left[\frac{Q_q m_D A_{b1}}{p_P\cdot p_{\gamma}}+\frac{2 Q_q E_{b2} m_P^3}{\left(p_P\cdot p_{\gamma}\right)^2}\right]\left[\left(p_P{\cdot}p_{\gamma}\right)\varepsilon _{\gamma}^{{\mu}*}-\left(p_P{\cdot}\varepsilon\right)p_{\gamma}^{\mu}\right]
\right.\\
&\left.+2\left(Q_q\frac{p_P{\cdot}\varepsilon}{p_P{\cdot}p_{\gamma}}\right)A_{a2}m_P p_P^{\mu}\right\},\\
&\mathcal{A}_c =\frac{- i e G_F V_{Qq}}{2\sqrt{2}}\left(u_l P_{\text {L$\mu $}}v_{\bar{\nu}}\right)\left\{ \frac{m_PA_{a2}}{p_l{\cdot}p_{\gamma}}i \epsilon ^{\alpha \beta \mu \sigma}p_{\gamma \beta }\varepsilon^* _{\gamma \alpha}p_{P\sigma}\right.\\
&\left.+\frac{m_P A_{a2}}{p_l\cdot p_{\gamma}}\left[\left(p_P{\cdot}p_{\gamma}\right)\varepsilon _{\gamma}^{{\mu}*}-\left(p_P{\cdot}\varepsilon\right)p_{\gamma}^{\mu}\right]
\right.\\ &\left.
+\frac{p_l{\cdot}\varepsilon}{p_l{\cdot}p_{\gamma}}A_{a2}m_P p_P^{\mu}\right\}
\label{eq:add1}.
\end{split}
\end{equation}
The above equations show that the contribution of each diagram in Fig. \ref{fig:feynsd} is not gauge invariant separately, but the sum of them is indeed gauge invariant, which is given in the following
\begin{equation}
\begin{split}
\mathcal{A}_{a+b+c} =& \frac{- i e G_F V_{Qq}}{2\sqrt{2}}\left\{i V\epsilon ^{\alpha \beta \mu \sigma}p_{\gamma \beta }\varepsilon^*_{\gamma \alpha}p_{P\sigma}
\right.\\ &\left.
+A\left[\left(p_P{\cdot}p_{\gamma}\right)\varepsilon _{\gamma}^{{\mu}*}-\left(p_P{\cdot}\varepsilon\right)p_{\gamma}^{\mu}\right]
\right.\\ &\left.
+2\left[\left(Q_q-Q_Q\right)\frac{p_P{\cdot}\varepsilon}{p_P{\cdot}p_{\gamma}}+\frac{p_l{\cdot}\varepsilon}{p_l{\cdot}p_{\gamma}}\right]A_{a2}m_D p_P^{\mu}\right\}\\
&\times \left(u_l P_{\text {L$\mu $}}v_{\bar{\nu}}\right).
\label{eq:12}
\end{split}
\end{equation}
This equation clearly shows that the sum of the contributions of all the diagrams in Fig. \ref{fig:feynsd} is
gauge invariant.  In eq.(\ref{eq:12})  the factors $V$ and $A$ are
\begin{equation}
\begin{split}
V=&-\frac{Q_Q m_P A_{a1}}{p_P{\cdot}p_{\gamma}}-\frac{Q_q m_P A_{b1}}{p_P{\cdot}p_{\gamma}}+\frac{m_P A_{a2}}{p_l{\cdot}p_{\gamma}}\\
&+\frac{2\left(Q_q F_{b2}-Q_Q F_{b1}\right)m_P^3}{\left(p_P{\cdot}p_{\gamma }\right)^2},\\
A=&-\frac{Q_Q m_P A_{a1}}{p_P\cdot p_{\gamma}}+\frac{Q_q m_P A_{b1}}{p_P{\cdot}p_{\gamma}}+\frac{m_P A_{a2}}{ p_l{\cdot}p_{\gamma}}\\
&+\frac{2\left(Q_q E_{b2}-Q_Q E_{b1}\right)m_P^3}{\left(p_P{\cdot}p_{\gamma }\right)^2}
\label{eq:13}.
\end{split}
\end{equation}
Next we shall calculate the coefficients $A_{a1}$, $A_{a2}$, $E_{a2}$ and $F_{a2}$.

The pseudoscalar meson state can be written in terms of the quark-antiquark creation and annihilation operators
\begin{equation}
\begin{split}
&{\mid}P\left(P=0\right)>= \frac{1}{\sqrt{3}}\sum _{\substack i}\int d^3k {\Psi _0}(|\vec{k}|)\\
& \times\frac{1}{\sqrt{2}}\left[b_Q^{i+}(\vec k,\uparrow)d_q^{i+}(-\vec k,\downarrow)-b_Q^{i+}(\vec k,\downarrow)d_q^{i+}(-\vec k,\uparrow)\right]\mid 0>,
\label{eq:14}
\end{split}
\end{equation}
where $i$ is the color index, the factor $1/\sqrt{3}$ the normalization factor for color indices, and $\vec{k}$ the 3-momentum of the quarks in the rest frame of the heavy meson. The wave function ${\Psi _0}(|\vec{k}|)$ has been calculated in the relativistic potential model previously, the numerical solution of the wave function can be fitted in the exponential form \cite{col,sdbase}
\begin{equation}
\Psi _0\left(|\vec{k}|\right)=4{\pi}{\sqrt{\lambda ^3 m_P}}e^{-\lambda |\vec{k}|}.
\label{eq:15}
\end{equation}
The numerical solutions of the parameter $\lambda$ for $D$, $D_s$ and $B$ mesons are quoted from Ref. \cite{sdbase} recently
\begin{displaymath}
\begin{array}{cc}
\lambda_D = 3.4{\rm GeV}^{-1},& \lambda_{D_s} = 3.2{\rm GeV}^{-1}, \\
\lambda_B=2.8{\rm GeV}^{-1}.&
\end{array}
\end{displaymath}
With the meson state given in eq. (\ref{eq:14}), the matrix element $<0\mid \bar{Q}\Gamma q\mid P>$ can be calculated straightforwardly, the result is
\begin{equation}
\begin{split}
&<0\mid \bar{Q}\Gamma q\mid P>=\sqrt{\frac{3}{2}}\frac{1}{8\pi^3}\int dkd{\Omega} k^2 {\Psi _0}\left(k\right)\\
&\times tr\left[M \cdot \Gamma \right]\sqrt{\frac{m_Q m_q}{p_Q^0 p_q^0}},
\label{eq:16}
\end{split}
\end{equation}
where $M=u_q(\vec k,\uparrow)\bar{v}_Q(-\vec k,\downarrow)-u_q(\vec k,\downarrow)\bar{v}_Q(-\vec k,\uparrow)$ is the Dirac spinner for the pseudoscalar meson, it can be obtained in the Dirac-Representation as
\begin{equation}
M=\frac{-\frac{1}{2}\left(\slashed p_q + m_q\right)\left(1+{\gamma}^0\right)\left(\slashed p_Q + m_Q\right){\gamma}^5}{2\sqrt{m_Q^0 m_q^0\left(p_Q^0+m_Q\right)\left(p_q^0+m_q\right)}}
\label{eq:17}.
\end{equation}
\subsection{\label{sec:level3}III The Long-Distance Contribution}

\begin{figure}
\includegraphics[scale=0.7]{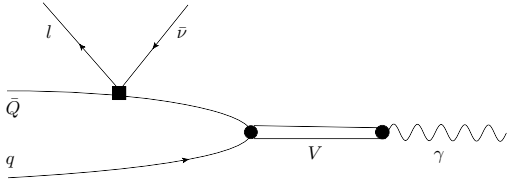}
\caption{\label{fig:ld} The LD contribution.}
\end{figure}

In this section, we estimate the contributions of long-distance physics. According to the spirt of vector meson dominance model \cite{VMD1,VMD2,VMD3,VMD4,VMD5}, we consider the resonance process $P^-\to l\bar{\nu} V\to l\bar{\nu}\gamma$, where the intermediate vector resonance $V$ can be $\rho$, $\omega$ and $\phi$. The contribution comes from the semileptonic intermediate $l\bar{\nu} V$, followed by the vector resonance turning to an on-shell photon $V\to \gamma$, which is shown in Fig.~\ref{fig:ld}. The amplitude of the long-distance contribution can be written as
\begin{equation}
\begin{split}
\mathcal{A}_{LD}=&\frac{eG_F V_{Qq}Q_q}{\sqrt{2}}u_l{\gamma}^{\mu}\left(1-{\gamma}_5\right)v_{\bar{\nu}}\\
&\times\sum _{\substack {V}}\varepsilon _{\gamma}^{\alpha*}<0\mid \bar{q}{\gamma ^{\alpha}}q\mid V>\frac{i e^{\phi _V}}{p_V^2-m_V^2+i m_V {\Gamma }_V}\\
&\times<V \mid \bar{Q}{\gamma}^{\mu}\left(1-{\gamma}_5\right)q\mid P >
\label{eq:19}.
\end{split}
\end{equation}
For the decays of $D$ and $B$ mesons, $V$ represents for $\rho$ and $\omega$ mesons, while for $D_s$ meson decay, $V$ represents for $\phi$ meson. $\phi _V$ is the relative phase between the long and short-distance contributions. The matrix element $<0 \mid \bar{q}{\gamma}^{\mu}q\mid V >$ is used to define the decay constant of the vector meson
\begin{equation}
<0 \mid \bar{q}{\gamma}^{\mu}q\mid V >=C_{qV}f_{V}\varepsilon _{V}^{\alpha *}
\label{eq:20},
\end{equation}
where the factor $C_{qV}$'s are
\begin{equation}
C_{u\rho}=C_{u\omega}=C_{d\omega}=\frac{1}{\sqrt{2}},C_{d\rho}=-\frac{1}{\sqrt{2}},C_{s\phi}=1
\label{eq:21}.
\end{equation}
The vector's decay constant $f_V$ can be derived from the decay rate of $V\to e^+ e^-$. After a short calculation, the decay constants can be related to the vector meson's leptonic decay widths
\begin{equation}
\begin{split}
&f_{\rho}^2=\frac{3 m_{\rho}^3}{4 \pi {\alpha}^2}\frac{2}{\left(Q_u-Q_d\right)^2}{\Gamma }_{\rho\to e^+ e^-},\\
&f_{\omega}^2=\frac{3 m_{\omega}^3}{4 \pi {\alpha}^2}\frac{2}{\left(Q_u+Q_d\right)^2}{\Gamma }_{\omega\to e^+ e^-},\\
&f_{\phi}^2=\frac{3 m_{\phi}^3}{4 \pi {\alpha}^2}\frac{1}{Q_s^2}{\Gamma }_{\phi\to e^+ e^-},
\label{eq:22}
\end{split}
\end{equation}
where $\alpha$ is the electromagnetic fine structure constant, $Q_u$, $Q_d$ and $Q_s$ are the charges of the quarks $u$, $d$ and $s$, respectively.

The hadronic matrix element $<V \mid J^{\mu}\mid P >$ in eq.~(\ref{eq:19}) can be decomposed according to its Lorentz structure as \cite{bsw1,bsw2}
\begin{equation}
\begin{split}
&<V \mid J_{V-A\mu}\mid P >=\frac{2}{m_P+m_V}\epsilon _{\mu\nu\rho\sigma} \varepsilon _V^{\nu}p_P^{\rho}p_V^{\sigma}V\left(q^2\right)\\
&+i\left\{\varepsilon _{V\mu}\left(m_P+m_V\right)A_1\left(q^2\right)\right.\\
&\left.-\frac{\varepsilon _V\cdot q}{m_P+m_V}\left(p_P+p_V\right)_{\mu}A_2\left(q^2\right)\right.\\
&\left.-\frac{\varepsilon _V\cdot q}{q^2}2m_Pq_{\mu}A_3\left(q^2\right)\right\}
+i\frac{\varepsilon _V\cdot q}{q^2}2m_Vq_{\mu}A_0\left(q^2\right),
\label{eq:23}
\end{split}
\end{equation}
where $q=p_P-p_V$, and $V$, $A_0$, $A_1$, $A_2$ and $A_3$ are form factors.

To obtain the amplitude gauge invariant, we take the trick used in Ref.\cite{dht} in treating the long-distance contribution in the precess $b\to s{\gamma}$ via the resonance $J/\Psi$. With the Lorentz decomposition of the hadronic matrix element in eq.(\ref{eq:23}), the product of the two matrix elements in eq.(\ref{eq:19}) can be calculated to be
\begin{equation}
\begin{split}
&\varepsilon _{\gamma\alpha}^*<0 \mid \bar{q}{\gamma}^{\alpha}q\mid V ><V \mid J_{V-A}^{\mu}\mid P >=\\
&e Q_q C_{qV}f_V \left\{\frac{2}{m_P+m_V}\epsilon^{\alpha\beta\mu\sigma}p_{P\sigma}p_{V\beta}\varepsilon _{\gamma\alpha}^*V\left(q^2\right)\right.\\
&\left.-iA_1\left(q^2\right)\left[\varepsilon _{\gamma}^{\mu *}\left(m_P+m_V\right)-\frac{\left(m_P+m_V\right)\left(p_P{\cdot}\varepsilon _{\gamma}\right)}{p_P{\cdot}p_V}p_V^{\mu}\right]\right.\\
&\left.-i\left(p_P{\cdot}\varepsilon _{\gamma}\right)\left[\frac{\left(m_P+m_V\right)p_V^{\mu}}{p_P{\cdot}p_V}A_1\left(q^2\right)\right.\right.\\
&\left.\left.-\frac{\left(p_P+p_V\right)^{\mu}}{m_P+m_V}A_2\left(q^2\right)+\frac{2 m_V q^{\mu}}{q^2}\left(A_3\left(q^2\right)-A_0\left(q^2\right)\right)\right]\right\}
\label{eq:24}.
\end{split}
\end{equation}
In the rest-frame of the heavy meson $P$, the product of the four-momentum of the meson $P$ and the polarization vector of the photon satifies $p_P{\cdot}\varepsilon = 0$. Then the last term in the above equation can be dropped. With $p_V=p_{\gamma}$, we obtaine
\begin{equation}
\begin{split}
\mathcal{A}_{LD}=&\sum _{\substack {V}}\frac{e G_F V_{Qq} Q_q}{\sqrt{2}}u_l{\gamma}^{\mu}\left(1-{\gamma}_5\right)v_{\bar{\nu}}\\
&\times \left\{i V_{LD}^V \epsilon _{\alpha \beta \mu \sigma}p_{\gamma }^{\beta }\varepsilon _{\gamma }^{{\alpha}*}p_P^{\sigma}
\right.\\ &\left.
+A_{LD}^V\left[\left(p_P{\cdot}p_{\gamma}\right)\varepsilon _{\gamma}^{{\mu}*}-\left(p_P{\cdot}\varepsilon\right)p_{\gamma}^{\mu}\right]\right\},
\label{eq:25}
\end{split}
\end{equation}
where:
\begin{equation}
\begin{split}
&V_{LD}^V=\frac{e^{i\phi _V}C_{qV}f_V}{-m_V^2+i m_V {\Gamma}_V}\frac{2}{m_P+m_V}V(q^2)\\
&A_{LD}^V=\frac{e^{i\phi _V}C_{qV}f_V}{-m_V^2+i m_V {\Gamma}_V}\frac{m_P+m_V}{p_P{\cdot}p_{\gamma}}A_1(q^2)
\label{eq:26}.
\end{split}
\end{equation}

\subsection{\label{sec:level4}IV Numerical Result and Discussion}

The numerical calculation is performed in the center-of-mass frame of the heavy meson, and the momentum of the photon is taken as $\left(E_{\gamma},0,0,-E_{\gamma}\right)$. For the input parameters, the masses of the quarks are taken as
\begin{displaymath}
\begin{array}{ll}
m_u = m_d=0.08{\rm GeV},& m_s = 0.30{\rm GeV}, \\
m_b = 4.98{\rm GeV},& m_c = 1.54{\rm GeV},
\end{array}
\end{displaymath}
which are taken to be consistent with that used to derive the wave function of the heavy meson in the relativistic potential model in Ref. \cite{sdbase}.

The electron and neutrinos are taken to be massless, the masses of the other leptons are taken from PDG \cite{particaldatagroup}.  The Cabibbo-Kobayashi-Maskawa (CKM) matrix elements are:
\begin{displaymath}
V_{cd}=0.2259, V_{cs}=0.974, V_{ub}=0.00389.\\
\end{displaymath}

\begin{figure}
\includegraphics[scale=0.35]{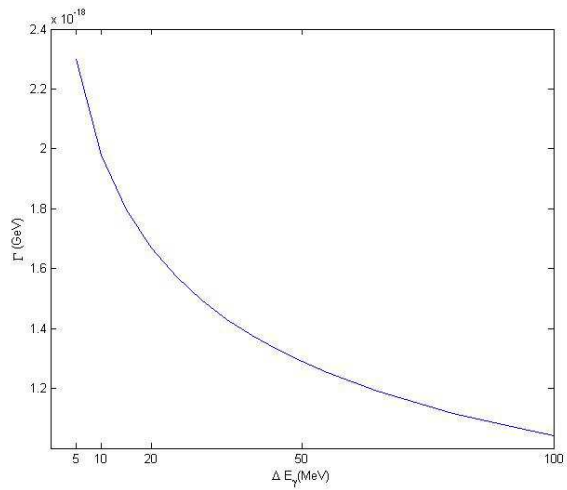}
\caption{\label{fig:irdiv} Decay widths of $D^-\to \gamma e \bar{\nu}$ as a function of $\Delta E_{\gamma}$.}
\end{figure}

The infrared divergence appears as the energy of the real photon goes to soft limit or the momentum of the photon is parallel to the momentum of a massless lepton. This divergence can be canceled when the decay width of the radiative leptonic decay is added with the pure leptonic decay width, in which one-loop radiative corrections are included \cite{irdiv}. The radiative leptonic decay with the energy of the photon lower than the experimental resolution can not be distinguished from the pure leptonic decay. Only photons with the energy larger than the experimental resolution can be distinguished. Therefore the radiative leptonic decay width depends on the photon energy resolution. The photon energy resolution can be a few MeV in experiment \cite{alice}. The dependence of the decay width on the resolution $\Delta E_{\gamma}$ is shown in Table \ref{tab:IRDIV} and Fig.~\ref{fig:irdiv}. For example, if taking $\Delta E_{\gamma}=10\;{\rm MeV}$, the decay width and branching ratio of $D\to \gamma e \bar{\nu}_e$ are
\begin{equation}
\begin{split}
&\Gamma (D\to \gamma e \bar{\nu}_e)=1.98\times 10^{-18}{\rm GeV},\\
&Br(D\to \gamma e \bar{\nu}_e)=3.29\times 10^{-6}
\label{eq:27}.
\end{split}
\end{equation}
In the following all the numerical calculation is performed by taking the resolution $\Delta E_{\gamma}=10\;{\rm MeV}$.

\begin{table}
\caption{\label{tab:IRDIV}The dependence of the decay width of $D^-\to \gamma e \bar{\nu}$ on the photon
resolution  $\Delta E_{\gamma}$.}
\begin{tabular}{cccc}\hline\hline
$\Delta E_{\gamma}$&$\Gamma _{D^-\to \gamma l \bar{\nu}}$&$\Delta E_{\gamma}$&$\Gamma _{D^-\to \gamma l \bar{\nu}}$\\
(MeV) & $(\times 10^{-18}{\rm GeV})$ & (MeV) & $(\times 10^{-18}{\rm GeV})$\\
\hline
$5 $ & $2.30$ & 55  & $1.25 $\\
$10$ & $1.98$ & 60  & $1.22 $\\
$15$ & $1.80$ & 65  & $1.19 $\\
$20$ & $1.67$ & 70  & $1.16 $\\
$25$ & $1.57$ & 75  & $1.14 $\\
$30$ & $1.49$ & 80  & $1.12 $\\
$35$ & $1.43$ & 85  & $1.10 $\\
$40$ & $1.38$ & 90  & $1.08 $\\
$45$ & $1.33$ & 95  & $1.06 $\\
$50$ & $1.29$ & 100 & $1.04 $\\ \hline
\end{tabular}
\end{table}

We also checked that the decay width is not sensitive to the cutoff of the angle between the momentum of the photon and the electron. We find taking different values for the angle cutoff $\Delta \theta$, the result changes extremely slowly.

To show the contribution of each diagram in Fig.\ref{fig:feynsd}, we list each diagram's contribution to the decay width in Table \ref{tab:SDResult}. It is shown that, although the amplitude of the diagram with the photon emitted from the heavy quark is suppressed by a factor of the heavy quark mass in the dominator of the propagator, which is like $\frac{i}{\slashed p_{\gamma}-\slashed p_Q - m_Q}$, for the cases of $D$ and $D_s$ mesons, the mass of $c$ quark is not large enough, so the contributions of diagrams $a$, $b$ and $c$ are all at the same order, but for the case of $B$ decay, the suppression is large, the contribution of diagram $b$ dominates. It can also be shown that, the contributions of the diagrams  in Fig.\ref{fig:feynsd} interfere destructively, especially in the case of $D$ and $D_s$ mesons, this is consistent with Ref.  \cite{workbefore2}.
\begin{table}
\caption{\label{tab:SDResult} The decay width contributed by each diagram in Fig. \ref{fig:feynsd}. In the first column, $\Gamma_a$, $\Gamma_b$ and $\Gamma_c$ means the contributions of diagrams $a$, $b$ and $c$, respectively, $\Gamma_{a+b}$ the contribution of the diagrams $a$ and $b$, $\Gamma_{a+b+c}$ the total contribution of diagrams $a$, $b$ and $c$.}
\begin{tabular}{cccc}\hline\hline
Width&$D^-\to \gamma e \bar{\nu}_e$&$D_s^-\to \gamma e \bar{\nu}_e$&$B^-\to \gamma e \bar{\nu}_e$ \\
(GeV) & & &  \\
\hline
$\Gamma_a$ & $1.37\times 10^{-17}$ & $3.06\times 10^{-16}$ & $9.14\times 10^{-21}$\\
$\Gamma_b$ & $1.35\times 10^{-17}$ & $1.94\times 10^{-16}$ & $1.01\times 10^{-18}$\\
$\Gamma_c$ & $9.78\times 10^{-18}$ & $3.23\times 10^{-16}$ & $5.83\times 10^{-20}$\\
$\Gamma_{a+b}$ & $9.86\times 10^{-18}$ & $3.08\times 10^{-16}$ & $0.96\times 10^{-18}$\\
$\Gamma_{a+b+c}$ & $1.98\times 10^{-18}$ & $8.46\times 10^{-18}$ & $0.82\times 10^{-18}$\\ \hline
\end{tabular}
\end{table}

We present the branching ratios with only the short-distance contributions for all the decay modes in TABLE.\ref{tab:allSDResult}. The branching ratio for $D^-\to \gamma \tau \bar{\nu}_{\tau}$ is very small, because the mass of $\tau$ is very large, the phase space for this decay mode is too small.
\begin{table}
\caption{\label{tab:allSDResult}Branching ratios of radiative leptonic decays of $B$ and $D$ mesons.
$BR_{SD}$ is the branching ratios with only the short-distance contribution. $BR_{LD}$ stands for branching ratio with only long-distance contribution. For $D$ and $B$ mesons,
$BR_{LD1}$ is the branching ratios of long-distance contribution via $\rho$ meson, while
for $D_s$ meson, the long-distance contribution is via $\phi$ meson. $BR_{LD2}$ is the branching ratios
of long-distance contribution via $\omega$ meson.}
\begin{tabular}{cccc}\hline\hline
Modes&$BR_{SD}$&$BR_{LD1}$ &$BR_{LD2}$\\
\hline
$D^-\to \gamma e \bar{\nu}_e$ & $3.3\times 10^{-6}$ & $7.5\times 10^{-6}$ & $6.3\times 10^{-6}$\\
$D^-\to \gamma \mu \bar{\nu}_{\mu}$ & $1.6\times 10^{-5}$ & $7.3\times 10^{-6}$ & $6.1\times 10^{-6}$\\
$D^-\to \gamma \tau \bar{\nu}_{\tau}$ & $5.5\times 10^{-9}$ & $9.1\times 10^{-10}$ & $7.6\times 10^{-10}$\\
\hline
$D_s^-\to \gamma e \bar{\nu}_e$ & $6.8\times 10^{-6}$  & $1.0\times 10^{-4}$ & -\\
$D_s^-\to \gamma \mu \bar{\nu}_{\mu}$ & $2.0\times 10^{-4}$ & $1.0\times 10^{-4}$ & -\\
$D_s^-\to \gamma \tau \bar{\nu}_{\tau}$ & $1.1\times 10^{-6}$ & $6.5\times 10^{-8}$ & -\\
\hline
$B^-\to \gamma e \bar{\nu}_e$ & $2.1\times 10^{-6}$  & $5.8\times 10^{-7}$ & $4.0\times 10^{-7}$\\
$B^-\to \gamma \mu \bar{\nu}_{\mu}$ & $2.0\times 10^{-6}$ & $5.8\times 10^{-7}$ & $4.0\times 10^{-7}$\\
$B^-\to \gamma \tau \bar{\nu}_{\tau}$ & $1.6\times 10^{-6}$ & $3.8\times 10^{-7}$ & $2.7\times 10^{-7}$\\
\hline
\end{tabular}
\end{table}

The short-distance branching ratios obtained in this work is slightly smaller than the previous works \cite{workbefore2}. More details about the quark momentum distribution are included in this work by using wave function obtained in the relativistic potential model.


To calculate the long-distance contribution, the transition amplitude $V\to {\gamma}$ is needed. The transition amplitude is related to the decay constant defined in eq.(\ref{eq:20}). Using the data on the decay rate of $V\to l^+l^-$ given in PDG \cite{particaldatagroup}, the decay constants of the vector mesons can be extracted as
\begin{equation}
\begin{split}
&f_{\rho}=0.169869{\rm GeV}, \\
&f_{\omega}=0.154631{\rm GeV},\\
&f_{\phi}=0.231784{\rm GeV}.
\end{split}
\label{eq:28}
\end{equation}
The $q^2$ dependence of the form factors defined in the hadronic matrix element $<V|J^{\mu}_{V-A}|P>$ are taken to be the usual pole approximation as
\begin{displaymath}
V(q^2)=\frac{h_v}{1-\frac{q^2}{M_V^2}}, \;A_1(q^2)=\frac{h_{a1}}{1-\frac{q^2}{M_{A1}^2}}.
\end{displaymath}
The parameters in the form factors for $D\to V$ and $B \to V$ transitions are quoted from Refs. \cite{pball} and \cite{pball2}, they are
\begin{displaymath}
\begin{array}{ll}
h_{v_{D\to \rho}}=1.0,& h_{a1_{D\to \rho}}=0.5;\\
M_{V_{D\to\rho}}=2.5\;{\rm GeV};& M_{A1_{D\to\rho}}=2.5\;{\rm GeV}; \\
h_{v_{B\to \rho}}=0.323,& M_{V_{B\to\rho}}^2=38.34\;{\rm GeV}^2;\\
h_{a1_{B\to \rho}}=0.242,& M_{A1_{B\to\rho}}^2=37.51\;{\rm GeV}^2;\\
h_{v_{B\to\omega}}=0.293,& M_{V_{B\to \omega}}^2=37.45\;{\rm GeV}^2;\\
h_{a1_{B\to \omega}}=0.219,& M_{A1_{B\to\omega}}^2=37.01\;{\rm GeV}^2.
\end{array}
\end{displaymath}
For the form factor of $D\to \omega$, we assume it is the same as that of $D\to \rho$. The form factors of $D_s\to \phi$ is taken from \cite{dsform} as
\begin{displaymath}
h_{v_{D_s\to \phi}}=1.21,\; h_{a1_{D_s\to \phi}}=0.55,\; M_{V_{D_s \to \phi}}=2.08\;{\rm GeV}.
\end{displaymath}
The form factor $A_1$ for $D_s\to \phi$ transition is approximated as a constant because the $q^2$ dependence of $A1$ is very weak \cite{dsform}.

With the parameters given above, the long-distance contributions to the decay width for each decay mode can be estimated, they are listed in Table~\ref{tab:allSDResult}, where the short-distance and long-distance contributions to the branching ratios of each decay mode are presented separately .  It shows that for decays of $D$ and $D_s$ mesons, long-distance contributions are as large as or even larger than the short-distance contributions, while for the case of $B$ decays, short-distance contributions dominate, long-distance contributions are roughly $4\sim 5$ times smaller than short-distance contributions.

\begin{figure}
\includegraphics[scale=0.5]{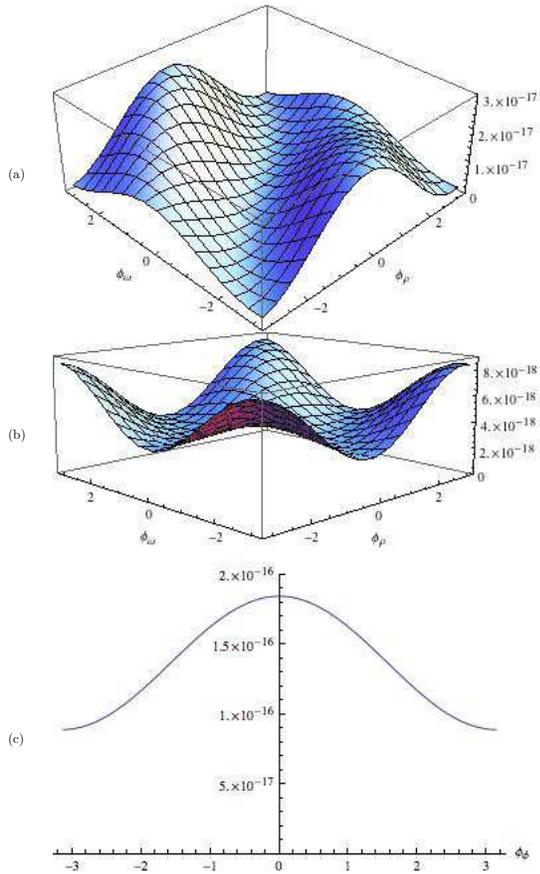}
\caption{\label{fig:allampl}From up to down, they are decay widths (${\rm GeV}$) of $D^-$, $B^-$ and $D_s^-\to\gamma e^- \bar{v}_e$ decays, including both the short and long-distance contributions, as functions of $\phi _{\rho}$ and $\phi _{\omega}$ in the cases of $D^-$ and $B^-$ meson decays, and $\phi _{\phi}$ in the case of $D_s^-$ meson decay. }.
\end{figure}

To get the total decay width, including both the short and long-distance contributions, one has to know the relative phase between the long and short-distance amplitudes.  Unfortunately we do not konw the relative phases exactly upto now. We have to leave the relative phases as free parameters.   To show how the decay width depends on the relative phases, we show the decay widths of $B$, $D$  and $D_s\to \gamma e\bar{\nu}_e$  decays in Fig.~\ref{fig:allampl} as an example. In the case of $D\to \gamma e\bar{\nu}_e$ decay, because the long-distance contributions are as important as the short-distance contributions, the relative phases between the long and short-diatance contributions can affect the decay widths considerably, the decay widths can change several times.  For $D_s\to \gamma e\bar{\nu}_e$ decay, the long-distance contribution dominates (see Table \ref{tab:allSDResult}), the dependence of the total branching ratio on the relative phase is weak.  While for $B\to\gamma e\bar{\nu}_e$ meson decay, the amplitudes of the short and long-distance contributions are at the same order, therefore the decay width still depends on the relative phase severely. For the case of the other decay modes, the dependence of the total branching ratios on the relative phases are illustrated numerically in Table \ref{tab:AllResult}. The situation is similar to the $B$, $D$ and $D_s\to\gamma e\bar{\nu}_e$ decay modes. The contribution of the long-distance physics is important, in general the branching ratio heavily depends on the relative phase between the long and short-distance contributions. Some decay modes can be greatly enhanced by the long-distance contributions. The branching ratio of $D_s\to\gamma e\bar{\nu}_e$ decay can be enhanced from the order $10^{-6}$ to $10^{-4}$.

\begin{table*}
\caption{\label{tab:AllResult}Total branching ratios of the radiative leptonic decays including both the short and
long-distance contributions. In case of $D$ meson decays, for illustration, the relative phase $\phi=\phi _{\omega}$ and
$\phi _{\rho}=0$ are taken. While for $B$ meson decays, $\phi=\phi _{\rho}=\phi _{\omega}$ is taken, and
in case of $D_s$ meson decays, $\phi$ is $\phi _{\phi}$.}
\begin{tabular}{ccccccc}\hline\hline
Modes& $BR_{tot}(\phi=0^o)$ & $BR_{tot}(\phi=30^o)$ & $BR_{tot}(\phi=60^o)$ & $BR_{tot}(\phi=90^o)$ & $BR_{tot}(\phi=120^o)$ & $BR_{tot}(\phi=150^o)$\\
\hline
$D\to \gamma e \bar{\nu }_e$ & $4.0\times 10^{-6}$ & $1.1\times 10^{-5}$ & $3.0\times 10^{-5}$ & $4.1\times 10^{-5}$ & $3.4\times 10^{-5}$ & $1.5\times 10^{-5}$\\
$D\to \gamma \mu \bar{\nu }_{\mu}$ &$2.5\times 10^{-5}$&$3.1\times 10^{-5}$&$4.8\times 10^{-5}$&$5.9\times 10^{-5}$&$5.2\times 10^{-5}$&$3.5\times 10^{-5}$\\
$D\to \gamma \tau \bar{\nu }_{\tau}$ &$5.7\times 10^{-9}$&$7.5\times 10^{-9}$&$1.2\times 10^{-8}$&$1.4\times 10^{-8}$&$1.2\times 10^{-8}$&$8.0\times 10^{-9}$\\
\hline
$B\to \gamma e \bar{\nu }_e$ &$6.7\times 10^{-6}$&$8.9\times 10^{-6}$&$1.3\times 10^{-5}$&$1.4\times 10^{-5}$&$1.2\times 10^{-5}$&$8.2\times 10^{-6}$\\
$B\to \gamma \mu \bar{\nu }_{\mu}$ &$6.8\times 10^{-6}$&$8.9\times 10^{-6}$&$1.3\times 10^{-5}$&$1.4\times 10^{-5}$&$1.2\times 10^{-5}$&$8.3\times 10^{-6}$\\
$B\to \gamma \tau \bar{\nu }_{\tau}$ &$5.3\times 10^{-6}$&$6.7\times 10^{-6}$&$9.2\times 10^{-6}$&$1.0\times 10^{-5}$&$8.8\times 10^{-6}$&$6.3\times 10^{-6}$\\
\hline
$D_s\to \gamma e \bar{\nu }_e$ &$1.5\times 10^{-4}$&$1.3\times 10^{-4}$&$0.9\times 10^{-4}$&$0.7\times 10^{-4}$&$0.9\times 10^{-4}$&$1.3\times 10^{-4}$\\
$D_s\to \gamma \mu \bar{\nu }_{\mu}$ &$4.5\times 10^{-4}$&$4.4\times 10^{-4}$&$4.1\times 10^{-4}$&$3.9\times 10^{-4}$&$4.1\times 10^{-4}$&$4.4\times 10^{-4}$\\
$D_s\to \gamma \tau \bar{\nu }_{\tau}$ &$1.0\times 10^{-6}$&$1.0\times 10^{-6}$&$1.2\times 10^{-6}$&$1.3\times 10^{-6}$&$1.2\times 10^{-6}$&$1.0\times 10^{-6}$\\ \hline
\end{tabular}
\end{table*}

\subsection{\label{sec:level5}V Summary}

The radiative leptonic decays of $B$, $D$ and $D_s \to l\bar{\nu}\gamma$ are studied in this work. The short-distance contribution is calculated by using the wave functions of the heavy mesons obtained in the relativistic potential model, more details about the quark-momentum distribution are included in this work. In addition to the short-distance contribution, the long-distance contribution is also estimated in the vector meson dominance model. The study shows that the long-distance contributions can seriously affect the decay rates. The branching ratio of $D_s^-\to \gamma e \bar{\nu_e}$ can be enhanced to the order of $10^{-4}$, which should only be at the order of $10^{-6}$ if only considering the short-distance contribution.

This work is supported in part by the
National Natural Science Foundation of China under contracts Nos.
10575108, 10975077, 10735080, 11125525 and by the Fundamental Research Funds for the
Central Universities No. 65030021.

\end{document}